\documentclass{article}
\usepackage{spconf,amsmath,graphicx,hyperref,booktabs}
\usepackage{marginnote}
\usepackage{setspace} 
\usepackage{comment}
\usepackage{multicol}
\usepackage{multirow}
\usepackage[table]{xcolor}
\usepackage{collcell} 
\usepackage{xcolor}
\usepackage{colortbl}

\title{Quantifying Articulatory Coordination as a Biomarker for Schizophrenia}
\name{Gowtham Premananth \& Carol Espy-Wilson \thanks{This work was supported by the National Science Foundation grant numbered 2124270.}}
\address{Institute for System Research, Department of Electrical and Computer Engineering,\\  
University of Maryland, College Park, USA\\
}
\definecolor{darkred}{RGB}{230,30,30}
\definecolor{lightgreens}{RGB}{150,250,150}
\definecolor{greens}{RGB}{0,210,0}
\begin{document}
%
\maketitle
\begin{abstract}
Advances in artificial intelligence (AI) and deep learning have improved diagnostic capabilities in healthcare, yet limited interpretability continues to hinder clinical adoption. Schizophrenia, a complex disorder with diverse symptoms including disorganized speech and social withdrawal, demands tools that capture symptom severity and provide clinically meaningful insights beyond binary diagnosis. Here, we present an interpretable framework that leverages articulatory speech features through eigenspectra difference plots and a weighted sum with exponential decay (WSED) to quantify vocal tract coordination. Eigenspectra plots effectively distinguished complex from simpler coordination patterns, and WSED scores reliably separated these groups, with ambiguity confined to a narrow range near zero. Importantly, WSED scores correlated not only with overall BPRS severity but also with the balance between positive and negative symptoms, reflecting more complex coordination in subjects with pronounced positive symptoms and the opposite trend for stronger negative symptoms. This approach offers a transparent, severity-sensitive biomarker for schizophrenia, advancing the potential for clinically interpretable speech-based assessment tools.
\end{abstract}


\section{Introduction}
\vspace{-0.3cm}

The rapid progress of artificial intelligence (AI) and the advent of large deep learning models have demonstrated impressive performance improvements across diverse domains \cite{thayyib2023state}, including healthcare applications \cite{bohr2020rise}. However, while many of these models deliver high diagnostic accuracy, their lack of interpretability remains a critical barrier to real-world clinical adoption \cite{amann2022explain}. For deployment in clinical practice, predictive systems must not only produce reliable results but also provide interpretable insights that clinicians can trust and act upon. This limitation has sparked growing interest in the development of interpretable AI models for healthcare, particularly in discovering clinically meaningful biomarkers that link computational predictions to underlying biological and behavioral phenomena \cite{ng2023benefits}.

Schizophrenia is a complex and debilitating mental health disorder that affects millions of individuals worldwide \cite{institute2021global}. Its heterogeneous presentation, encompassing symptoms such as hallucinations, disorganized thought and speech as well as negative symptoms like social withdrawal, creates major challenges for accurate diagnosis and effective treatment. Although automated approaches for detecting schizophrenia have shown promise \cite{premananth24_interspeech}, simply identifying the disorder is insufficient to meaningfully support clinical decision-making. Recent research has therefore shifted toward automated assessment of symptom severity and subtypes of schizophrenia, with the goal of enabling more personalized interventions and assisting clinicians in prioritizing assessment and care \cite{premananth25_interspeech}.

Within this context, speech has emerged as a particularly promising modality for mental health assessment. Automated speech-based methods have been successfully applied to disorders such as depression \cite{seneviratne2022multimodal} and schizophrenia \cite{speech_schizo}, but most existing studies rely on self-supervised speech representations \cite{wav2vec}. While effective, these features provide little interpretability, making it difficult to identify which aspects of speech drive the diagnostic predictions. Researchers address this limitation by leveraging articulatory features, which capture how speech sounds are produced and offer greater transparency and clearer clinical insights. Articulatory features not only capture meaningful patterns of disordered speech but have also shown potential in supporting schizophrenia diagnosis and severity estimation \cite{speech_schizo}. Building on this motivation, this study seeks to develop interpretable biomarkers derived from speech-based articulatory coordination features to improve both the transparency and clinical utility of automated schizophrenia assessment systems.

\section{Dataset}
\vspace{-0.3cm}

The dataset used in this study was collected at the University of Maryland School of Medicine in collaboration with the University of Maryland, College Park, as part of a mental health research initiative \cite{Kelly2020-cj}. It includes participants diagnosed with schizophrenia, depression, and healthy controls, all of whom took part in multiple in-clinic interview sessions that were recorded. Before each of these sessions the subjects were evaluated using symptomatology questionnaires like the Brief Psychiatric Rating Scale (BPRS) \cite{overall1962brief} to assess the severity of different symptoms they were exhibiting at the time of their session. For the experiments reported in this paper, we used a subset of the dataset containing only subjects with schizophrenia and healthy controls. The subset consists of a total of 140 sessions belonging to 39 unique subjects whose BPRS scores varied between a range of 19 to 62.

\section{Data Pre-processing \& Feature Extraction}
\vspace{-0.3cm}
The audio recordings in the dataset contained speech from both the interviewer and the subject. To ensure subject-specific analysis, the recordings were first diarized and segmented into 40-second intervals, which were then used for feature extraction. Articulatory features were extracted from each speech segment using an acoustic-to-articulatory speech inversion system \cite{speechinversion}. Specifically, six Vocal Tract Variables (TVs) were estimated: lip aperture, lip protrusion, tongue tip constriction degree, tongue tip constriction location, tongue body constriction degree, and tongue body constriction location. In addition, articulatory source features related to aperiodicity and periodicity were obtained through an Aperiodicity Periodicity Pitch (APP) detector \cite{appdetector}. 

\section{Methodology}
\vspace{-0.3cm}

The extracted TVs were combined with the extracted source features. We further computed the Full Vocal Tract Coordination (FVTC) structure to capture the phasing relationships between articulatory gestures. This was achieved using a channel-delay correlation mechanism \cite{FVTC}, which calculates both auto-correlation and cross-correlation of the vocal tract variables across multiple delay scales. As previous studies that used the full vocal tract coordination as inputs for deep learning models has shown promising results for schizophrenia classification\cite{embc} and schizophrenia symptoms severity estimation \cite{premananth25_interspeech} we chose to investigate on how this full vocal tract coordination can be effectively used as an interpretable bio marker for schizophrenia detection.

As the next step, Eigenspectra were obtained from correlation matrices (Full Vocal tract Coordination Matrices) through eigenvalue decomposition, with eigenvalues arranged in descending order of magnitude. These rank ordered eigenspectra were then used to analyze patterns of articulatory coordination in speech, following the framework proposed by Seneviratne et al. \cite{c1}. In their study, eigenspectra derived from speech feature-based correlation matrices were employed to differentiate between healthy individuals and those with depression. This approach demonstrated how articulatory coordination features can serve as markers of underlying speech dynamics, providing a quantitative method for assessing coordination in complex vocal behaviors.

To characterize differences between speech-based gestural coordination, Seneviratne et al.\cite{c1} simulated three distinct types of signal combinations designed to resemble different coordination trends observed in speech: overly simplified speech (a group of sine waves with phase shifts limited to 0 and 180 degrees), natural speech (sine waves with phase shifts of 0, 90, and 180 degrees), and erratic speech (sine waves with random phase shifts). Coordination between these signal combinations was calculated and then using eigenvalue decomposition, rank ordered eigenspectra was obtained from these coordination matrices. Analysis of the resulting eigenspectra from these simulated sine wave groups revealed clear distinctions among their coordination patterns. This analysis was performed by calculating the difference between the generated eigenspectra under 2 scenarios: difference between overly simplified and natural speech, and difference between erratic and natural speech. Simplified speech's eigenspectra difference produced eigenvalue distributions where lower-ranked values started at high positive magnitudes, dipped into the negative range, and stabilized near zero. Erratic speech's eigenspectra difference, in contrast, displayed the opposite trajectory, beginning with very high negative values, rising to a positive peak, and gradually decreasing toward zero. These findings underscore the value of analysis done on eigenspectra obtained from coordination matrices in capturing subtle but informative differences in the temporal organization of speech gestures, offering insights into both natural and disordered speech. 

We applied the same methodology to compare individuals diagnosed with schizophrenia with healthy controls. For each 40-second segment in the dataset, we computed the FVTC matrices and extracted their corresponding rank ordered eigenspectra. The eigenspectra from all healthy control segments were then averaged to generate a generalized reference eigenspectrum. For the schizophrenia group, the eigenspectra were averaged within each subject across their segments to obtain subject-level eigenspectra. Finally, we generated individual difference spectrum for each subject by calculating the difference between the subject-level eigenspectra for each subject from the schizophrenia group and the generalized healthy control reference eigenspectrum.


To quantify the rank-ordered difference spectrum, we employed a weighted sum with an exponential decay factor. Given a ranked spectrum $v$ with a length of $n$ and a decay factor $\alpha$ the weighted sum with an exponential decay (WSED) is calculated as follows,

\begin{equation}
    WSED = \sum_{i=1}^{n}{v_{i}.\alpha^{i-1}}
\end{equation}
    
In this approach, higher-ranked components contribute more prominently to the final measure, while the influence of lower-ranked components diminishes progressively according to the decay rate. This weighting strategy ensures that the most informative dimensions of the eigenspectrum are emphasized, while still retaining contributions from the full spectrum of components. By applying exponential decay, the method balances sensitivity to dominant features with robustness against noise from lower-ranked dimensions, thereby providing a more interpretable and stable quantification of the eigenspectrum's structure.

WSED scores were first computed for each individual 40-second speech segment from their corresponding eigenspectra differences with an $\alpha=0.8$. Session-level scores were then derived by averaging the WSED values across all segments within a session. Finally, subject-level scores were obtained by averaging the session-level scores corresponding to each subject. All the WSED scores are normalized to keep them within the range of -1 to +1.

\section{Results and Discussion}
\vspace{-0.3cm}

\begin{figure}[h!]
\includegraphics[width=0.9\linewidth]{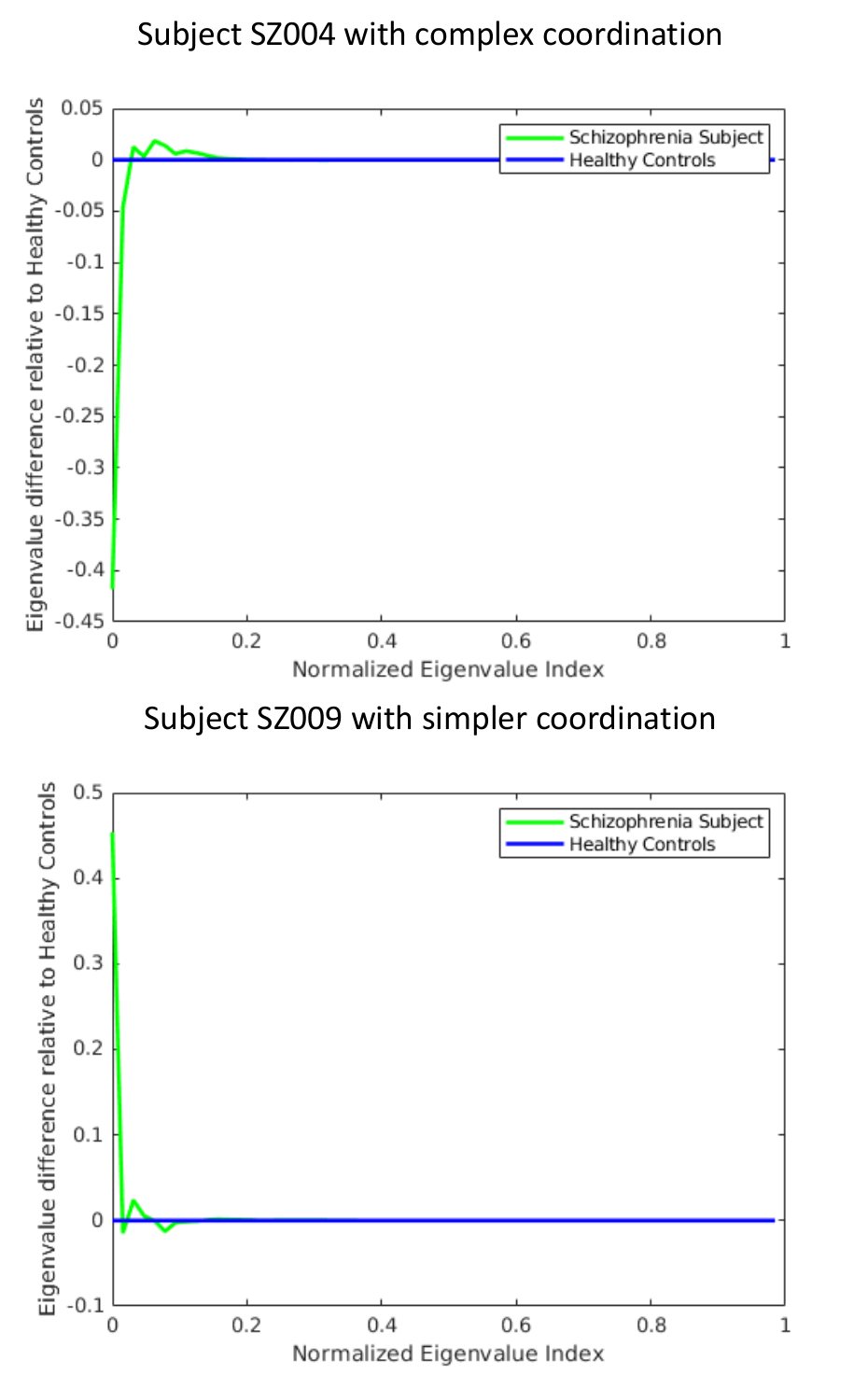}
\caption{ Eigenspectra difference plots of a subject with complex corrdination and a subject with a more simpler coordination}
\label{fig:eigen}
\end{figure}
\vspace{-0.2cm}

In the initial set of experiments, eigenspectra difference plots were generated for all subjects with schizophrenia. Separate plots were created for each individual, revealing that 13 out of the 23 subjects exhibited more complex, erratic coordination patterns, while the remaining subjects displayed simpler coordination. An example of an eigenspectra difference plot with a complex coordination and an example with a simpler coordination is shown in Fig.\ref{fig:eigen}. This discrepancy raised an important question: why didn't all individuals with schizophrenia demonstrate similar coordination trends in their speech? Therefore we looked into it more deeply by focusing on the severity of the subjects and the quantified values of the generated eigenspectra difference plots.

\begin{table}[h!]
\begin{center} 
\caption{\textbf{Schizophrenia subjects with their subject-wise WSED scores and eigenspectra difference plot trends} (The WSED scores are color coded with lowest values starting from red and transitioning to green when the scores increase)} 
\label{score_table}
\vspace{0.25cm}
\begin{tabular}{|l |c |c|} 
\hline
\multirow{2}{*}{SubjectID} & \multirow{2}{*}{WSED scores} & Eigenspectra Difference\\
&&Plot's Trend\\
\hline
SZ001	&\cellcolor{red!20!yellow}-0.0040&complex\\
\hline
SZ002	&\cellcolor{red}-0.0990 &complex\\
\hline
SZ004	&\cellcolor{darkred}-0.2489&complex\\
\hline
SZ005	&\cellcolor{yellow!10!lightgreens}0.1626&simple\\
\hline
SZ008	&\cellcolor{greens!90!lightgreens}0.2990&simple\\
\hline
SZ009	&\cellcolor{greens}0.3433&simple\\
\hline
SZ010	&\cellcolor{lightgreens!85!greens}0.1925&simple\\
\hline
SZ014	&\cellcolor{yellow!20!lightgreens}0.1369&simple\\
\hline
SZ015	&\cellcolor{lightgreens!30!greens}0.2473&simple\\
\hline
SZ016	&\cellcolor{yellow!40!lightgreens}0.1248&simple\\
\hline
SZ019	&\cellcolor{yellow}0.0505&complex\\
\hline
SZ020	&\cellcolor{red!60!yellow}-0.0170&complex\\
\hline
SZ022	&\cellcolor{yellow!50.0!lightgreens}0.0970&complex\\
\hline
SZ024	&\cellcolor{red!80!yellow}-0.0503&complex\\
\hline
SZ025	&\cellcolor{darkred!80!red}-0.1937&complex\\
\hline
SZ026	&\cellcolor{yellow!30!lightgreens}0.1286&complex\\
\hline
SZ027	&\cellcolor{lightgreens!50!greens}0.2076&simple\\
\hline
SZ028	&\cellcolor{darkred!40!red}-0.1618&complex\\
\hline
SZ033	&\cellcolor{red!40!yellow}-0.0166&complex\\
\hline
SZ037	&\cellcolor{darkred!20!red}-0.1156&complex\\
\hline
SZ042	&\cellcolor{lightgreens!40!greens}0.2195&simple\\
\hline
SZ049	&\cellcolor{lightgreens!70!greens}0.1890&simple\\
\hline
SZ056	&\cellcolor{darkred!60!red}-0.1872&complex\\
\hline
\end{tabular} 
\end{center} 
\end{table}
\vspace{-0.5cm}
The results of the WSED score calculations for subjects with schizophrenia are presented in Table \ref{score_table}, alongside the corresponding eigenspectra difference plot trends (complex or simple). Among the 23 schizophrenia subjects analyzed, those exhibiting complex coordination consistently obtained negative WSED scores, while those with simpler coordination received positive scores. The only exceptions were three subjects (SZ001, SZ019, and SZ022), whose scores were very close to zero (yellowish colors in the table as they are near the median which is 0). When all subjects’ scores were ranked, these three fell near the exact midpoint (10th, 11th, and 12th positions), indicating that ambiguity arises only in a very narrow region around zero. Overall, these findings suggest that WSED scores obtained from eigenspectra differences provide a reliable and effective means of quantifying the coordination patterns in speech production.

\begin{figure}[h!]
\includegraphics[width=0.9\linewidth]{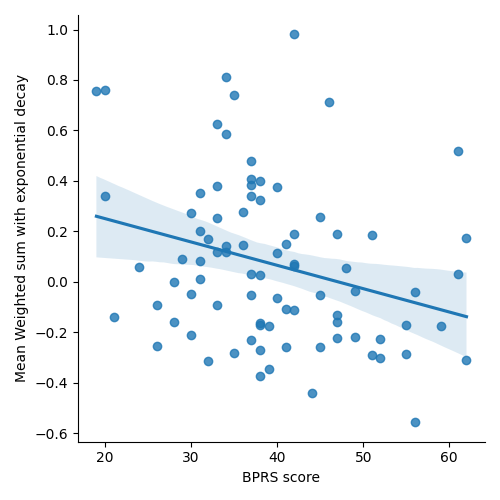}
\caption{Mean WSED plotted in relation with the BPRS scores of the Schizophrenia subjects}
\label{fig:WSED_BPRS}
\end{figure}
\vspace{-0.2cm}

Furthermore, we examined the relationship between the calculated WSED scores from the eigenspectra difference plots and the overall severity of schizophrenia symptoms. To explore this, WSED scores were plotted against the Brief Psychiatric Rating Scale (BPRS) scores for each session, as shown in Fig.\ref{fig:WSED_BPRS}. The plot reveals a clear trend: sessions with lower BPRS scores tend to correspond to higher WSED scores, indicating simpler coordination patterns, whereas sessions with higher BPRS scores are associated with lower WSED scores, reflecting more complex coordination.

\begin{figure}[h!]
\includegraphics[width=0.9\linewidth]{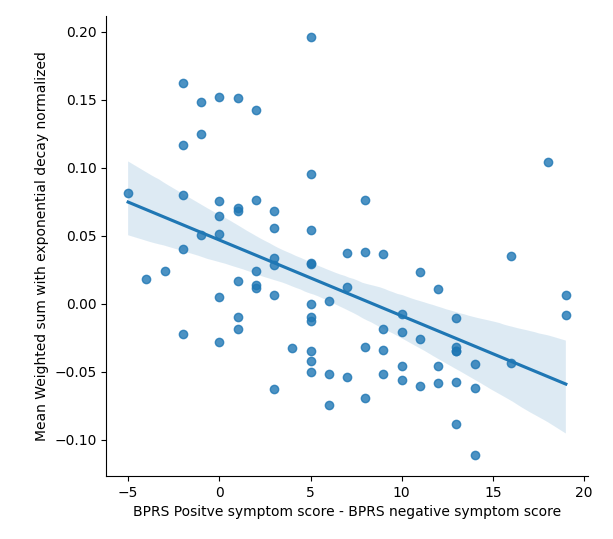}
\caption{Mean WSED plotted in relation with the Difference of positive and negative symptoms subscale scores}
\label{fig:WSED_POS_NEG}
\end{figure}

Given the complexity of schizophrenia and its diverse symptom profiles, we further examined the relationship between WSED scores and the difference between the Positive and Negative symptom subscales of the BPRS (Fig.\ref{fig:WSED_POS_NEG}). The results show a clear trend: subjects with more pronounced positive symptoms exhibited lower WSED scores, indicating more complex coordination, whereas those with stronger negative symptoms showed the opposite pattern.

\section{Conclusion and Future Work}
\vspace{-0.3cm}

In this study, we introduced an interpretable framework for assessing schizophrenia severity using eigenspectra difference plots and weighted sum with exponential decay (WSED) scores derived from articulatory speech features. Eigenspectra difference plots distinguished between complex and simpler coordination patterns, with 13 out of 23 individuals exhibiting more erratic coordination. The WSED metric further quantified these patterns, consistently separating subjects with complex coordination (negative scores) from those with simpler coordination (positive scores). Only a small subset of subjects showed scores near zero, indicating that ambiguity is largely confined to a narrow boundary region. Importantly, WSED scores demonstrated a strong relationship with overall BPRS symptom severity, underscoring their clinical relevance. Beyond this result, analysis of the Positive and Negative BPRS subscales revealed a meaningful trend: subjects with more pronounced positive symptoms tended to have lower WSED scores, reflecting more complex coordination, while those with stronger negative symptoms showed the opposite pattern. Together, these findings highlight the potential of WSED scores as clinically meaningful biomarkers and reinforce the promise of interpretable articulatory-based features in moving beyond binary diagnosis toward nuanced, severity-sensitive assessment of schizophrenia.

Looking ahead, several avenues for future work can enhance and extend this framework. First, expanding the dataset to include a larger and more diverse population would improve the generalizability and robustness of the findings. Second, longitudinal analyses could explore how WSED scores evolve over time and whether they can serve as indicators for treatment response or disease progression. Third, integrating additional modalities such as video and text and corresponding feature representations may provide a richer multimodal characterization of schizophrenia symptoms. Finally, translating these methods into real-time clinical tools, with visualization strategies that maintain interpretability for clinicians, will be crucial for advancing adoption in healthcare settings. By addressing these directions, future research can further bridge the gap between computational speech analysis and practical, personalized mental health care.


\bibliographystyle{IEEEbib}
\bibliography{refs}

\end{document}